# Multi-stage domain adversarial style reconstruction for cytopathological image stain normalization

Xihao Chen[#], *Student Member*, *IEEE*, Jingya Yu[#], *Student Member*, *IEEE,* Li Chen, Shaoqun Zeng,

Xiuli Liu[*], Shenghua Cheng[*]

*Abstract*—The different stain styles of cytopathological images have a negative effect on the generalization ability of automated image analysis algorithms. This article proposes a new framework that normalizes the stain style for cytopathological images through a stain removal module and a multi-stage domain adversarial style reconstruction module. We convert colorful images into grayscale images with a color-encoding mask. Using the mask, reconstructed images retain their basic color without red and blue mixing, which is important for cytopathological image interpretation. The style reconstruction module consists of per-pixel regression with intra-domain adversarial learning, inter-domain adversarial learning, and optional task-based refining. Per-pixel regression with intra-domain adversarial learning establishes the generative network from the decolorized input to the reconstructed output. The inter-domain adversarial learning further reduces the difference in stain style. The generation network can be optimized by combining it with the task network. Experimental results show that the proposed techniques help to optimize the generation network. The average accuracy increases from 75.41% to 84.79% after the intra-domain adversarial learning, and to 87.00% after inter-domain adversarial learning. Under the guidance of the task network, the average accuracy rate reaches 89.58%. The proposed method achieves unsupervised stain normalization of cytopathological images, while preserving the cell structure, texture structure, and cell color properties of the image. This method overcomes the problem of generalizing the task models between different stain styles of cytopathological images.

## I. Introduction

THE discipline of medicine relies on inductive logic, empirical learning, and evidence-based usage. In recent years, artificial intelligence has played an increasingly important role in medical applications. In the field of cytopathology, the accumulation of digital pathological slide images provides a huge database for cytopathological image analysis, and artificial intelligence is increasingly used in the field of cytopathological image analysis, especially cervical slide images [1]. Assisted screening systems based on big data and artificial intelligence reduce subjective errors in manual interpretation while decreasing doctors' workloads and labor costs [2]. This is of great significance in popularizing cytopathology screening and reducing the incidence of cervical cancer.

However, the actual process of algorithm-based screening systems faces great challenges in terms of model generalization for different cytopathological images. There are differences in the image styles because of variations in pathological slide staining methods, staining reagent concentrations, scanning instruments, and instrument parameters [3]. Humans have good knowledge transfer ability, enabling high interpretation accuracy to be maintained across different stain styles, but models based on big data are sensitive to changes in the numerical distribution caused by the different stain styles. For example, a model trained on one stain style will struggle to achieve the same or similar performance with another style.

Automated image analysis models require better generalization ability to adapt to data from different staining styles. There has been some research on this problem [4] [5]. Converting RGB color maps to grayscale images is a simple way to increase the adaptability of the models. The texture feature of cells or tissues exist in the grayscale images, but most cases need to be judged by combining color information. Augmenting the training set through some data transformation and/or adding noise to the data can improve the generalization ability of the models, but the new data style is always limited. Data enhancement cannot guarantee that the model will perform well on any stain style [6]. Mixed training and transfer learning are effective methods for improving the performance of the model on new stain-style data [7], but require manual labeling of the new stain style, which is resource-intensive. Therefore, it would be highly beneficial to normalize the stain of cytopathological images under unsupervised conditions. By

* X. Liu and S. Cheng both are corresponding authors.
\# X. Chen and J. Yu contributed equally to this work.
X. Chen, J. Yu, S. Zeng, X. Liu and S. Cheng are with Britton Chance Center for Biomedical Photonics, Wuhan National Laboratory for Optoelectronics-Huazhong University of Science and Technology, Wuhan, Hubei 430074, China and MoE Key Laboratory for Biomedical Photonics, School of Engineering Sciences, Huazhong University of Science and Technology,

Wuhan, Hubei 430074, China. (email: melon@hust.edu.cn; jingyayu@hust.edu.cn; sqzeng@mail.hust.edu.cn; goodlixu@gmail.com; chengshen@hust.edu.cn)

L. Chen is with Department of Clinical Laboratory, Tongji Hospital, Huazhong University of Science and Technology, Wuhan, Hubei 430030, China. (email: chenliisme@126.com)

normalizing the stain of different styles of data, not only would the generalization ability of the models be improved, but the large-scale annotation of new styles would also be avoided. Therefore, unsupervised image stain normalization represents a very practical solution [8].

There are many studies on stain normalization. The traditional methods can be roughly divided into color separation methods [9] [4] [11] [12] [13] [14] and color modification methods [10] [15] [16] [17] [18] [20]. These methods usually refer to the target style images as source images and the images to be transferred as target images. Color separation methods decompose the image into its main components (usually indicating the used stain), then process the decomposed main components before integrating the processed components to obtain a final normalized image. Template color matching methods mainly perform a mathematical transformation on the target image so that it matches the source image. However, color separation methods ignore the spatial dependence between organizational structures, and thus struggle when the difference in stain types is large. Color modification methods rely on the selection of reference images and require a sufficient amount of reference image data [19]. In addition to traditional methods, some machine learning techniques for stain normalization have emerged over the past decade. In particular, those based on generative adversarial networks (GAN) tend to perform better than traditional methods [19] [20] [21] [22] [23].

To address the challenges outlined above, we propose a new unsupervised stain normalization framework consisting of a stain removal module and a multi-stage domain adversarial style reconstruction module. We convert the colorful images into grayscale images using a color-encoding mask. With this mask, reconstructed images can retain basic color without red and blue mixing, which is important for cytopathological image interpretation. The style reconstruction module consists of per-pixel regression with intra-domain adversarial learning, inter-domain adversarial learning, and optional task-based refinement. Per-pixel regression with intra-domain adversarial learning establishes the generative network from the decolorized input to the reconstructed output. Inter-domain adversarial learning between the reconstructed target domain and the source domain further reduces the difference in the stain style. At the same time, we optimize the generation network by combining it with the task network. Experimental results show that all of the proposed techniques are beneficial to the optimization of the generation network. The average accuracy of the task network on the source domain image prior to stain normalization is 75.41%. After the first stage, the intra-domain adversarial learning, the average accuracy increases to 84.79%. After further inter-domain adversarial learning, the average accuracy rate rises to 87.00%. After the third stage, which uses the guidance of the task network, the average accuracy rate increases to 89.58%.

The method proposed in this paper achieves unsupervised stain normalization of cytopathological images while preserving the cell structure, texture structure, and cell color properties of the image. This method solves the challenge of generalizing the task model between different stain styles of cytopathological images.

## II. RELATED WORK

Previous studies on stain normalization can be divided into the following three categories: color separation methods [4] [9] [11] [12] [13] [14], color modification methods [10] [15] [16] [17] [18] [20], and machine learning-based stain normalization [19] [20] [21] [22] [23]. A detailed description of these categories is given in the following subsections.

### A. Color Separation

The color separation methods decompose the source image and the target image into the main stains constituting the image, and then mathematically analyze the decomposed images to find a mapping relationship between the two. The respective decompositions of the target image are then reconstructed by a convolution operation, resulting in the reconstructed image having a similar homologous image dyeing style, thereby achieving the stain normalization of the target image.

A commonly used decomposition technique is the color deconvolution method. Color deconvolution is widely used in histopathological image analysis. First proposed by Ruifrok and Johnston [9], this method is a colorful image stain separation and quantification method. The color deconvolution matrix is determined by decompressing the color information collected by the RGB camera and using a color deconvolution vector for different stains. To separate the contribution of multiple stains in a sample based on specific RGB absorption and the Lambert–Beer law [24], the tissue is first stained separately, quantifying its corresponding stain absorption eigenvector, and then the contribution of the color of each stain to the final pixel value is linearly estimated. Although Ruifrok and Johnston provide standard stain matrices for various stain combinations, it is best to manually quantify the absorbance eigenvectors of a single stain because of the variability of actual situations, although this process is time consuming and labor intensive than existing approaches.

Khan et al. [14] proposed a nonlinear mapping method to standardize the staining and color classification processes for color matrix estimation. Each stain point is pixel-classified using a pre-trained classification model to generate a probability map. Probability plots are used to define the average color of each stain. When using different tissues, stains, or imaging tools, their methods have robust deconvolution matrix estimates and appropriate mapping functions result in less image artifacts than other approaches[10] [11].

The key to the color separation method is to estimate the eigenvectors of different stains, but this ignores the spatial information of the images and the dependence between different tissues. When three or more stains appear, or the source and target domains vary widely, the normalization effect of this method may not be especially good. Moreover, these methods are generally based on Lambert's reflectance model [25], whereas optical microscopy images are formed by light transmission of the sample. In addition, differences in the staining methods used in cytopathology and histopathology mean that there are differences in the eigenvectors of each stain. Thus, the widely used color deconvolution in histopathological images cannot be directly applied to cytopathological images.

### B. Color Modification

The color modification method was first proposed by

Reinhard et al. [10]. After converting the source image and the target image from the RGB space to the Lab space, they calculate the mean and standard deviation of the source image and the target image for each sub-channel, and then convert the target image into a picture based on the source image style through a simple mathematical transformation. This method of matching the cumulative histogram of the source image and the target image is operationally simple and works well for some similar natural images. However, in the processing of pathological images, this method ignores the inherent multi-modality of the data. When the selected target image and the source image are not balanced in terms of tissue composition, the background region may be mapped to the color region and the foreground region is incorrectly mapped.

To overcome this limitation of the template color matching method, the image can be segmented and color modification applied to different segmentation types. Reinhard et al. [10] proposed a solution based on manual segmentation, Magee et al. [15] used a probabilistic Gaussian mixture model for pixel categories, and Janowczyk [16] proposed the StaNoSA technique, which uses a sparse self-encoder to segment images into tissue subtypes so that each individual structure can perform color normalization independently. Principal component analysis was employed to characterize the color features based on a standard set of training data.

Color modification may introduce certain artifacts, and the image is often divided into different parts by segmentation. This causes the default pixel ratio of each dye type in the source image and the target image to be the same, which is obviously not always right. Although Kothari et al. [17] attempted to normalize the stain for all pixels equally instead of using frequency weighting, this resulted in noise or pixels of low-frequency colors being considered as important as the main color pixels.

*C. Machine Learning-based Stain Normalization*

Vahadane et al. [21] used sparse nonnegative matrix factorization to establish the basis of color normalization in an unsupervised manner in their structure retention color normalization (SRCN) method. SRCN is flexible and preserves the biological structure and stain density of the target image, and adopts adaptive patch samples instead of the whole image. However, the effect of SRCN relies on the color basis, which can lead to artifacts if determined incorrectly.

In view of the good performance of GANs for transferring styles, a number of GAN-based stain transfer or normalization methods have been developed [19] [22] [23]. There are three key issues for GAN-based stain transfer or normalization: how to ensure the generative target domain images are consistent in style with the source domain images, how to preserve the tissue structure and fine cell texture, and whether to combine the tasks.

Bentaieb et al. [19] developed a generator and a task-specific discriminator, and used an edge-weighted regularization loss to preserve the tissue structures. The task network and discriminator share the feature extraction module, which makes the generator better suited to task networks.

Cho et al. [22] proposed a stain transfer network composed of a decolorization function and a style generator. They ensured good stain-style transfer and preserved the tissue structure and fine cell texture by minimizing the L2-distance between the source image and the reconstructed image. Their model calculates a feature-preserving loss in terms of the features extracted from the task network in addition to the GAN loss, which allows the generator to adapt to the task network. However, the decolorization function simply converts a colorful image into a grayscale image. The process is linear and does not achieve true normalization of the difference in gray intensity. Therefore, the reconstructed source images and the reconstructed target images may exhibit some differences in style.

Shaban et al. [23] used the method of unpaired image-to-image translation recommended by CycleGAN [26] to achieve cycle stain transfer. The cycle-consistency in CycleGAN aims to preserve information such as the tissue structure and cell texture in the translation. This approach is independent of the specific analysis task, which results in moderate adaptation of the task network. In addition, the model needs to be completely retrained when the target domain changes.

The methods reviewed above have several advantages for stain normalization, but also have some specific disadvantages. In addition, most of them focus on histopathological images, and there is a lack of stain normalization methods for cytopathological images. In fact, in the field of cytopathological images, there are many cases where the image styles are inconsistent. Therefore, we propose a multi-stage domain adversarial style reconstruction for unsupervised cytopathological image stain normalization.

The proposed method consists of a stain removal module and a style reconstruction network. To ensure the style of the generative target domain images is consistent with that of the source domain images, we develop a multi-stage domain adversarial generative network for style reconstruction: per-pixel regression with intra-domain adversarial learning, inter-domain adversarial learning, and optional task-based refinement. Per-pixel regression with intra-domain adversarial learning establishes the generative network from the decolorized input to the reconstructed output. The inter-domain adversarial learning between the reconstructed target domain and the source domain further reduces the difference in the stain style. To preserve the cell texture and some useful color information, we convert the colorful images into grayscale images with a color-coding mask. From the grayscale images, our method can reconstruct the source stain style without loss of cell texture information through per-pixel regression and multi-stage adversarial learning. Using the mask, the reconstructed images retain their basic color without red and blue mixing, which is important for cytopathological image interpretation.

Experiments show that the proposed techniques are beneficial to the optimization of the generated network. When only the first two stages are applied (i.e., per-pixel regression with intra-domain adversarial learning and inter-domain adversarial learning), the stain normalization method is independent of the tasks and is suitable for any task. The third stage (i.e., task-based refinement) can optionally be used to further improve the reconstruction network for a specific task. The proposed method achieves unsupervised normalization of the stain style while retaining the image cell structure, staining texture structure, and cell color properties, thus solving the challenge of generalizing the task model between different stain styles of cytopathological images.

## III. METHOD

We present a model for unsupervised stain normalization through stain removal and multi-stage domain adversarial style reconstruction. The reconstruction model not only ensures the visual consistency of the cytopathological images after reconstruction of the stain style, but also realizes unsupervised domain adaptation, thus improving the accuracy of unlabeled cytopathological image analysis.

### A. Problem Setting

To better explain our proposed method, we assume that the source domain $\{X^S\}$ is the dataset with annotation $\{Y^S\}$ and the target domain $\{X^T\}$ is the unlabeled dataset. Differences in imaging instruments, instrument parameters, and pathological slide staining methods result in $\{X^S\}$ and $\{X^T\}$ presenting different staining styles, which is why we call them different domains. Our aim is to normalize the target domain style to the source domain style. The proposed model consists of four sub-networks, namely a style reconstruction network G, intra-domain discriminator D1, inter-domain discriminator D2, and task network T. The task network is trained based on the source domain with a large number of annotations, so it has excellent analysis performance in the source domain, but performs poorly in the target domain because of the different stain styles. The parameters of each subnetwork are assumed to be $\theta_G$, $\theta_{D1}$, $\theta_{D2}$, and $\theta_T$. In addition, we refer to the style-reconstructed image of a cytopathological image $x$ as $\hat{x}$.

### B. Model Definition

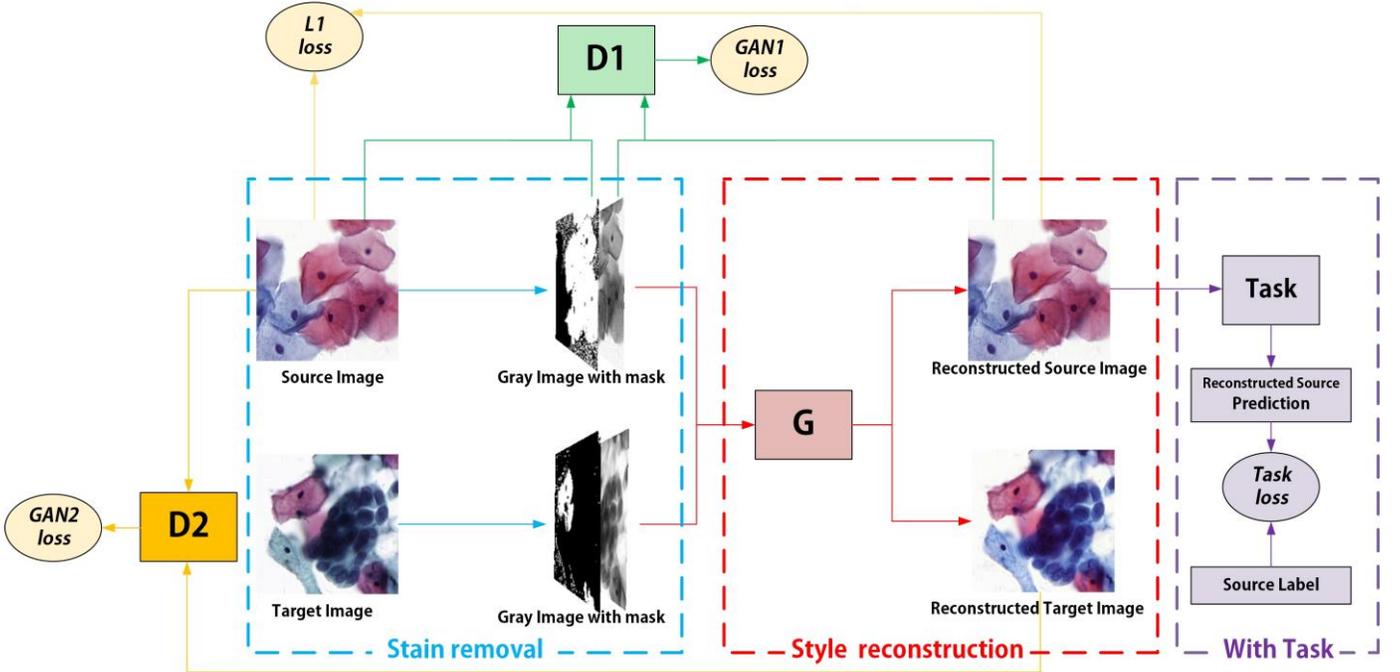

Fig. 1. Staining style reconstruction process with multi-stage domain adversarial style learning and task-based refinement. The proposed model normalizes the target image to the source image by stain removal and style reconstruction. During the process of style reconstruction, we take multiple losses to ensure that the reconstructed style is consistent with the source stain style, including GAN1 Loss (green), GAN2 Loss (orange), L1 Loss (yellow), and Task Loss (purple).

*1) Stain Removal Module*

The most intuitive discrepancy between the distributions of the source domain and the target domain is that the staining style is quite different. For the staining style reconstruction network G, we aim to ensure that: (i) G has the same performance for $\{X^S\}$ and $\{X^T\}$, meaning that it can reconstruct almost identical staining styles for cytopathological images from different domains; (ii) G ensures that the semantic information of the image representation is consistent before and after this process. In essence, G is similar to other convolutional neural networks and has a strong sensitivity to differences in numerical distributions. Only when G obtains an input image with a consistent distribution can the input image be mapped to the output image with the same distribution. To bind the input image to a nearly uniform distribution while preserving the original semantic information of the cytopathological images to the greatest extent, the stain removal module converts the original image into a two-channel image formed by superimposing a grayscale image and a color-encoding mask (called grayscale image with mask, GM) as an input to the style reconstruction network G, and provides G with the morphological information and rough color information of the original image.

*a) Grayscale Image*

The grayscale image is obtained by the weighted addition of the R, G, B channels of the original colorful image. This method preserves the image texture information while erasing the color difference of the colorful image [27]. In cytopathological images, besides erasing the stain style, the corresponding grayscale image ensures that G acquires cell morphology information.

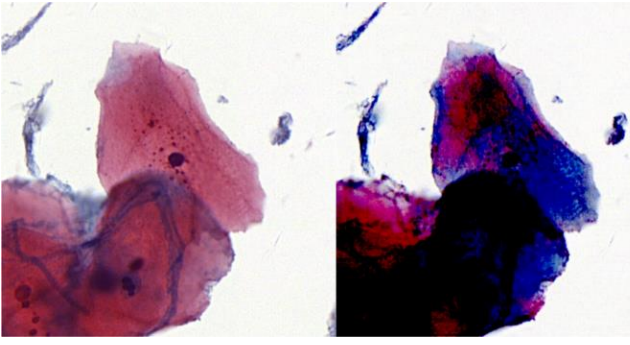

Fig. 2. Grayscale image only. Simply using a grayscale image as the input to G causes a chaotic red and blue distribution of the output image.

*b)    Color-Encoding Mask*

For cytopathological images, the color information contained in the stain style also contains a certain degree of semantic information. In the process of producing a cytopathological image, the staining reagent dyes the cell cytoplasm to red or blue according to the difference in acidity and alkalinity, and it is necessary to use such color information for manual interpretation or automatic interpretation with a task network. However, if G only obtains the grayscale image, the stain style reconstructed by G will ignore this information, resulting in a distribution problem with the red and blue in the cytoplasm (see Fig. 2). To solve the above problem, we convert the original cytopathological image into a color-encoding mask as part of the input to G, thus providing the rough color information of the cytoplasm. The specific color-encoding mask generation process is as follows: the natural image can be shown by the three channels of the RGB color space. We encode the pixel with the greatest value of the R channel as 1 and encode the others as 0.

*2) Multi-stage Domain Adversarial Learning*

As the task network T is trained on a large number of samples with annotations in the source domain, the model performance can be improved by choosing the source domain $\{X^S\}$ as the reconstruction style of G (called the source stain style). This means that the goal of training G is to make the reconstructed $\{\hat{X}^S\}$, $\{\hat{X}^T\}$ have a stain style that is as consistent as possible with $\{X^S\}$. To this end, we use multi-stage domain adversarial learning to train G.

*a)    Intra-Domain Adversarial Learning and Per-Pixel Regression*

The stain removal module can preserve fine morphology information of the cytopathological image while erasing the two domain's stain styles to obtain a similar distribution. Consequently, the distribution of $\{GM^S\}$ and $\{GM^T\}$ is roughly the same at this point. This means that, when we are able to restore $\{GM^S\}$ to $\{X^S\}$, we can also rebuild $\{GM^T\}$ to $\{X^S\}$.

In the adversarial training, G attempts to recover a more realistic $\hat{X}^S$ according to the input $GM^S$ to fool the intra-domain discriminator D1, whereas D1 wants to distinguish the reconstructed image $\hat{X}^S$ from the real image $X^S$, thus forming a process of adversarial training between G and D1 [28]. Naturally, this intra-domain ($\hat{X}^S$ and $X^S$ are both from the source domain) adversarial loss should be defined as follows:

$$L_{GAN1}(G, D1) = E_{X^S}[logD1(X^S)] + E_{GM^S}[1 - logD1(G(GM^S))] \quad (1)$$

In addition, as $GM^S$ is obtained directly by removing the stain component of $X^S$, we can use the per-pixel loss term to train G to achieve pixel-by-pixel regression. Following [29], we adopt the L1 distance loss, which gives the pixel-by-pixel loss while encouraging less blur rather than higher-order loss. The specific loss function is:

$$L_{L1}(G) = E_{X^S,GM^S}[\|X^S - G(GM^S)\|_1] \quad (2)$$

*b)    Inter-Domain Adversarial Learning*

In addition to color information such as hue, the images contain brightness, contrast, and intensity details. Although the stain removal module can normalize the input image of G to some extent by erasing the stain style, i.e., it can remove the difference in information carried by hue, differences between the other types of information persist in the grayscale image calculated linearly from the RGB color space, causing differences between the distribution of $\{GM^S\}$ and $\{GM^T\}$. Therefore, for $\{GM^T\}$, the stain style reconstructed by G, which is only trained with two losses, may struggle to completely match the source stain style $\{X^S\}$.

In the adversarial learning of GAN, the discriminator can perceive the difference between the generated image and the real image, thus encouraging the generator to generate a more realistic image. In other words, the generator actively erases the difference perceived by the discriminator. Inspired by this, we introduce a new form of adversarial training between two domains, called inter-domain adversarial training. In this loss, G attempts to reconstruct a more realistic $\hat{X}^T$ to fool the inter-domain discriminator D2, while D2 hopes to distinguish the reconstructed image $\hat{X}^T$ from the real image $X^S$, creating a new dynamic balance and forming a new adversarial learning. Similar to (1), the loss function of the inter-domain adversarial learning ($\hat{X}^T$ and $X^S$ are from different domains) is:

$$L_{GAN2}(G, D2) = E_{X^S}[logD2(X^S)] + E_{GM^T}[1 - logD2(G(GM^T))] \quad (3)$$

*3) Task-based Refinement*

For the GM converted from the source domain or target domain image through the stain removal module, the style reconstruction network G obtained by the multi-stage domain adversarial learning achieves fairly good reconstruction performance. However, $\{\hat{X}^S\}$ and $\{\hat{X}^T\}$ still have subtle differences from $\{X^S\}$, which will result in the task network exhibiting suboptimal performance on the reconstructed images. Thus, we propose to add a loss to the overall model to supervise the reconstruction results given by the task network, called $L_{Task}$ [30]. For the labeled source domain, $L_{Task}$ emphasizes that the distribution change caused by the stain style removal and reconstruction processes will not excessively change the prediction performance of the task network on the reconstructed style. Therefore, this task loss can be defined in classification form as the following cross-entropy loss:

$$L_{Task}(G) = E_{X^S,C^S}[-T(X^S)logT(G(C^S)) - (1 - T(X^S))(1 - logT(G(C^S)))] \quad (4)$$

As the task network has been fully trained based on the source domain, we replace $T(X^S)$ in (4) with the label $Y^S$ of $X^S$ in practical operations.

G adapts to the optimal distribution of the task network. Specifically, we fix the task network's parameter $\theta_T$ and adjust $\theta_G$ through the supervision information provided by $L_{Task}$.

Thus, the stain style that G can reconstruct is not only as close as possible to $\{X^S\}$, but also has a distribution that is better suited to the task network.

### C. Objectives

The goal of our proposed model is to reconstruct a stain style that is visually consistent with the source domain stain style through the style reconstruction network G, while offering excellent task prediction accuracy for the task network. Accordingly, the goals of the overall model can be defined as follows:

$$arg \min_{G,T} \max_{D1,D2} \lambda_{GAN1}L_{GAN1} + \lambda_{GAN2}L_{GAN2} + \lambda_{L1}L_{L1} + \lambda_{L_{Task}}L_{Task} \quad (5)$$

Here $\lambda_{GAN1}, \lambda_{GAN2}, \lambda_{L1}, \lambda_{L_{Task}}$ are hyperparameters indicating the relative importance of the different loss functions in the overall optimization process.

## IV. EXPERIMENTS

In this section, we quantitatively analyze the overall performance of the model through the task network's average prediction accuracy on the target domain before and after applying the proposed method. We also qualitatively evaluate the change in the target domain's stain style from the perspective of human vision or through the Bhattacharyya distance, and evaluate its consistency with the source stain style. Note that, although our task network is a classification model in the experiment, our recommended stain normalization model is equally effective for other areas of cytopathological image analysis, such as segmentation and detection.

### A. Experimental Dataset

We used two groups of cervical cytopathological slides (3D1 and 3D2) from Department of Clinical Laboratory, Tongji Hospital, Huazhong University of Science and Technology. The two groups were scanned by different 3DHISTECH slide scanners (3DHISTECH Ltd.). As these two groups of slides are different and were imaged by different scanners, they have very different stain styles and form a suitable dataset for testing our model. Group 3D1 was considered as the source domain and group 3D2 was the target domain. For both 3D1 and 3D2, we invited physiologists to label the lesion cells of positive slides. All cells of negative slides were considered to be normal cells. We used the annotations of the source domain to train the task network, and only used the annotations of the target domain to test the performance of our stain normalization model.

To train the task network of the source domain, we generated about 300000 positive patches by randomly sampling from 30950 annotations of lesion cells and about 300000 negative patches by randomly sampling from the negative slides. For the intra-domain and inter-domain adversarial training, $L_{L1}+L_{GAN1}+L_{GAN2}$, we used 50000 patches from the source domain and 50000 patches from the target domain. We used 8000 positive patches and 8000 negative patches from the target domain as the test dataset to evaluate the performance of our stain normalization method.

### B. Model Architecture

#### 1) Task Network

We used ResNet50 [31] as the basic network of the two-category network. The network input was the original colorful images of size $512 \times 512$. The training process record is shown in Fig. 3. Each block uses 3000 images as the training set and 500 images as the test set, with the ratio of positive data to negative data set to 1:1. The training process finally stabilized and the average accuracy of the training set and test set on the source domain was found to be 96.41% and 96.33%, respectively.

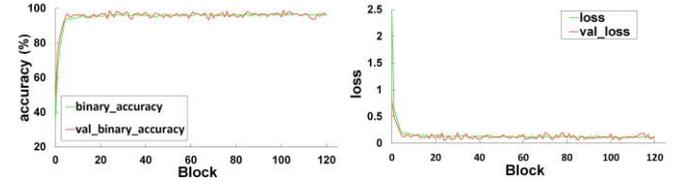

Fig. 3. Training record of the task network.

#### 2) Style Reconstruction Network G

For the style reconstruction network G, we used the U-net architecture with full convolution and skip connection [32]. For considerations of calculation speed and GPU memory, our G did not downsample the input image to a vector before upsampling, like the traditional U-net architecture. Instead, we adjusted the convolution stride size and downsampled the feature maps to 1/64 of the original image size through the six layers of the convolution operation, and then gradually upsampled to the original image size. Note that the convolution operation refers to sequential operation modules: convolution – batch normalization [33] – ReLU activation.

#### 3) Discriminator D1, D2

For the intra-domain discriminator D1 and inter-domain discriminator D2, we did not use a pooling layer to downsample. We used a five-layer convolution operation with strides to downsample the image and the flattening operation to turn the output of the final convolution layer into a vector [34]. Additionally, a fully connected layer was used to output the probability value. This convolution operation is similar to that of G, except that the activation function is the Leaky ReLU function.

### C. Experimental Design

We investigated the importance of each stage of our model. First, we verified the importance of stain information in the classification of cytopathological images. We then tested the performance degradation of the trained task networks in the target domain. To verify the superiority of our proposed reconstruction model and the importance of each module, we sequentially added each module to the experimental process and computed the task network's average prediction accuracy on the reconstructed images in the target domain.

(i) We used $L_{L1}$ and $L_{GAN1}$ to supervise the training of the style reconstruction network G ($L_{L1}+L_{GAN1}$);

(ii) On the basis of (i), we added $L_{Task}$ provided by the pretrained task network to supervise the training of G ($L_{L1}+L_{GAN1}+L_{Task}$);

(iii) On the basis of (i), we added $L_{GAN2}$ for supervision during the training of G ($L_{L1}+L_{GAN1}+L_{GAN2}$);

(iv) On the basis of (iii), we added $L_{Task}$ provided by the pretrained task network to supervise the training of G ($L_{L1}+L_{GAN1}+L_{GAN2}+L_{Task}$).

### D. Results

TABLE I
PERFORMANCE OF THE PROPOSED METHOD AND OTHER RELATED APPROACHES.

| Accuracy | %Mean±Variance | %Max |
|---|---|---|
| Color | 75.41±0.16 | 81.01 |
| Gray | 66.91±0.04 | 70.80 |
| Enhance_Alexnet | 56.32±0.01 | 58.09 |
| Enhance_our | 83.19±0.02 | 85.81 |
| $L_{L1}+L_{GAN1}$ | 84.79±0.04 | 87.68 |
| $L_{L1}+L_{GAN1}+L_{GAN2}$ | 87.00±0.04 | 90.22 |
| $L_{L1}+L_{GAN1}+L_{Task}$ | 86.90±0.02 | 89.53 |
| $L_{L1}+L_{GAN1}+L_{GAN2}+L_{Task}$ | 89.58±0.01 | 90.34 |

Mean and variance of the first four rows are calculated from the average accuracy of the top five task networks with the best performance on the unnormalized target images. Mean and variance of the fifth and sixth rows are calculated from the performance of the top five task networks on the optimally normalized target images. Methods in the last two rows are trained with the task network, so the mean and variance of the two are calculated from the performance of the combined task network on the target images with different normalization degrees. The maximum is the best performance of each method.

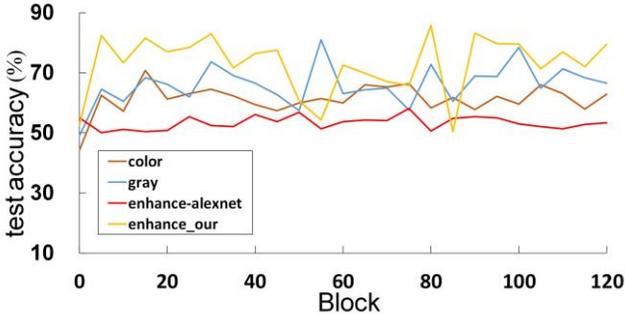

Fig. 4. Test accuracies on target images with the task networks trained on training datasets of different transformations. The brown line denotes the test accuracy on target images with the task network trained on the original source colorful images. The blue line denotes the test accuracy on gray target images with the task network trained on the gray source images. The red line denotes the test accuracy on the target images with the task network trained on the source images enhanced by PCA. The yellow line denotes the test accuracy on the target images with the task network trained on the source images enhanced by the proposed method.

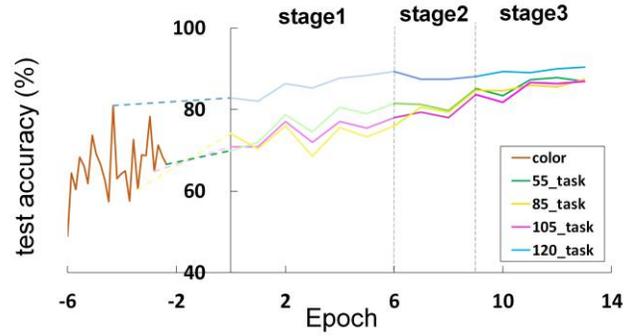

Fig. 5. Test accuracies on reconstructed target images with the task network. The brown curve is identical to that in Fig. 4 and denotes the test accuracy on target colorful images with the task network trained on the original source color images. We choose three task networks at different pre-trained stages (three points on the brown curve) and use them to test the average accuracy on the target images reconstructed by G at different stages. We train G by sequentially adding different losses. At stage 1, we only use $L_{L1}+L_{GAN1}$. We add $L_{GAN2}$ and $L_{Task}$ in stages 2 and 3, respectively.

*1) The Importance of Color Information*

Color information plays an important role in the automatic analysis of the model for cytopathological images. The average accuracy in the target domain of the task network trained on colorful and grayscale images from the source domain is 75.41% and 66.91%, respectively. In other words, when losing the color information, the average accuracy of the task network on the target domain drops by 8.50%.

*2) The Importance of Style Reconstruction*

(i) The difference in stain style between different domains seriously affects the prediction effect of the task network. This is because different stain styles give cytopathological images different distributions of intensity and hue. The task network trained using only one aspect does not adapt well to other distributions, inevitably leading to some degradation of the task network when applied to unfamiliar domains. For example, the average accuracy of the task network trained based on the source domain's colorful images is 96.33% on the source domain's test dataset, whereas the average accuracy on the target domain's test dataset is just 75.41%.

(ii) While G normalizes the stain styles of images from different domains, it can greatly improve the average accuracy of the task network prediction in the target domain. The reconstructed cytopathological images of the target domain are visually (Fig. 6) consistent with the source stain style, which significantly improves the ability of the task network to analyze the target domain's reconstructed images. As presented in Table I, the average accuracy of the task network on reconstructed target domain images by the complete style reconstruction G ($L_{L1}+L_{GAN1}+L_{GAN2}+L_{Task}$) reaches 89.58%, an increase of 14.17% over the target domain images without reconstruction.

(iii) Compared with the proposed method of stain normalization, the effect of traditional data augmentation methods is quite limited or unstable. The traditional PCA-based data enhancement method seems to be less effective for this task, with the average accuracy of the task network dropping to 56.32%. To this end, we designed a version of the data enhancement method for the cytopathological image dataset. Through the linear transformation of the HSV space, the average accuracy of the task network on the target domain test dataset increased to 83.19%.

*3) The Effect of Inter-domain Adversarial Learning*

In Table I, comparing $L_{L1}+L_{GAN1}$ and $L_{L1}+L_{GAN1}+L_{GAN2}$, it is clear that adding the inter-domain adversarial training $L_{GAN2}$ increases the average accuracy of the task network in the target domain (from 84.79% to 87.00%); in addition, $L_{L1} + L_{GAN1} + L_{Task}$ has a lower average accuracy than $L_{L1}+L_{GAN1}+L_{GAN2}+L_{Task}$. This confirms our earlier analysis: Through inter-domain adversarial training, G can effectively perceive the difference between $\{GM^S\}$ and $\{GM^T\}$, and reconstruct $\{\hat{X}^T\}$ to be more similar to $\{X^S\}$ by ignoring or making up for this difference. This greatly improves the average classification accuracy of the task network on $\{\hat{X}^T\}$.

*4) The Effect of Task-based Refinement*

We can verify the effect of the supervision information provided by $L_{Task}$ through two sets of comparisons: (i) Comparing $L_{L1} + L_{GAN1} + L_{Task}$ with $L_{L1} + L_{GAN1}$, the task network's average accuracy on the target domain increases by 2.11%; (ii) Comparing $L_{L1} + L_{GAN2} + L_{GAN2} + L_{Task}$ with $L_{L1}+L_{GAN2}+L_{GAN2}$, the task network's average accuracy on the target domain increases by 2.58%. When we provide supervision information regarding the task network to G, this implicitly requests G to reconstruct a style that is not only consistent with the source stain style, but also better suited to the task network. This improves the task network's performance on the reconstructed $\{\hat{X}^T\}$.

*5) Visual Consistency*

As can be seen from Fig. 5, the reconstructed target image is intuitively very similar to the source image. In addition, we can also observe subtle differences in the styles reconstructed by G trained by different loss function combinations. For example, the red in the style reconstructed under the supervision of $L_{L1} + L_{GAN1} + L_{GAN2}$ is more vivid than that supervised only by $L_{L1}+L_{GAN1}$, causing it to be closer to the red color in the source image. For $\{\hat{X}^S\}$, we can use its ground truth $\{X^S\}$ to measure the Bhattacharyya distance between their RGB histograms, which reflects the similarity between two probability distributions. The Bhattacharyya distances listed in Table II indicate that the reconstruction process supervised by $L_{L1}+L_{GAN1}$ reconstructs a similar stain style to that of the source. Additionally, when sequentially adding another two loss terms ($L_{GAN2}+L_{Task}$), the reconstruction effect is not significantly affected.

TABLE II.
BHATTACHARYYA DISTANCE OF $\{X^S\}$ AND $\{\hat{X}^S\}$

|  | Bhattacharyya distance |
|---|---|
| $L_{L1}+L_{GAN1}$ | 0.2067 |
| $L_{L1}+L_{GAN1}+L_{GAN2}$ | 0.2113 |
| $L_{L1}+L_{GAN1}+L_{GAN2}+L_{Task}$ | 0.2169 |

*6) Test Accuracy in Every Stage*

In Fig. 5, the three points denote the different accuracies of the three task networks on the original target images, including the optimal task network (blue), medium task network (yellow), and poor task network (purple). The recommended method greatly improves the analysis ability of all three task networks on the reconstructed target images. In addition, as the loss functions are sequentially added during the G training process, the similarity between the reconstructed target style and source style increases. This is confirmed by the test accuracy with the task network gradually rising as the training of G progresses (Fig. 5). Although simple preprocessing methods (such as enhancement) can increase the test accuracy of the task network on the target images (e.g., the optimal performance of "Enhance_our" reached 85.81%), the performance of the task network exhibits strong fluctuations and is unable to converge. In the training process of G, however, the performance of the task network on the target images reconstructed by G at different stages of training is relatively stable (Fig. 5).

### E. Model Sensitivity

The hyperparameters of the model were set to $\lambda_{GAN1}=1$, $\lambda_{GAN2}=1$, $\lambda_{L1}=100$, $\lambda_{L_{Task}}=10$. For the first two hyperparameters, we artificially set their values according to their specific roles. For $\lambda_{L1}$, the value of 100 was recommended by [29], because it results in less blur while ensuring the regression effect and the visual effect of the style reconstruction results. Through testing, we found that the model was insensitive to order of magnitude changes in $\lambda_{L_{Task}}$, which produced only a slight (0.5%) reduction in the average accuracy of the target domain reconstruction images.

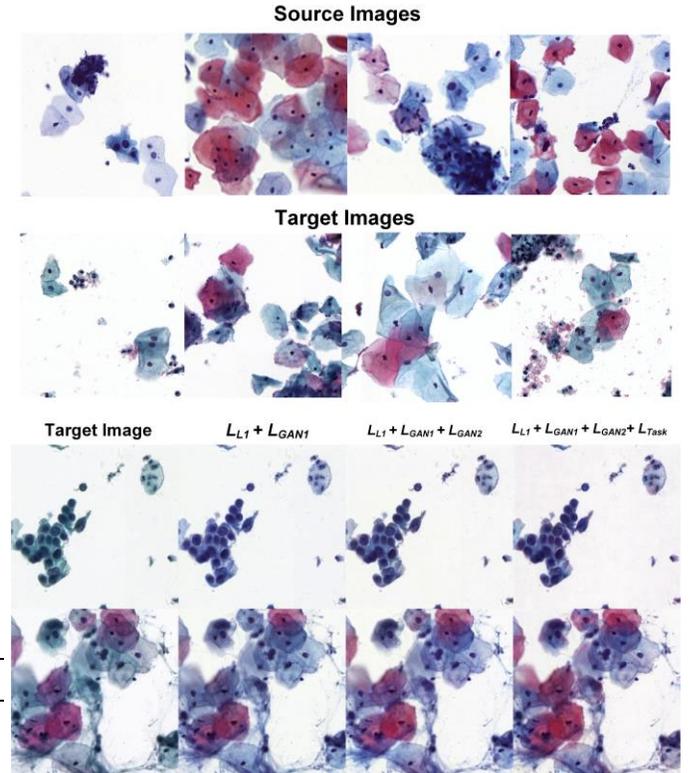

Fig. 6. Reconstruction results of the style reconstruction network G with different loss functions for supervised training.

### F. Model Recommendation Training Process

We trained G in stages. In the first stage of training, we only used $L_{GAN1}, L_{L1}$ to supervise the model. The goal was to enable G to learn a generative model from a decolorized input to the reconstructed output in $\{X^S\}$. In the second stage, we added $L_{GAN2}$ to adjust G so that it could be generalized between domains {X^S} and {X^T}. In the third stage, we further refined G with $L_{Task}$ to improve the generation of a stain style better suited to the task network.

## V. CONCLUSION

We have presented a new stain normalization framework for cytopathological images. The framework consists of a stain removal module and a multi-stage domain adversarial style reconstruction module. Our method can normalize images of different styles into a single style, while preserving the cell structure, texture structure, and cell color properties. This method overcomes the difficulty of generalizing task models between different stain styles of cytopathological images.

First, we transferred the colorful image into a gray image using a color-encoding mask in the stain removal module, which retains necessary information and ensures that the reconstructed image keeps its basic color without red and blue mixing. We then reconstructed the stain style according to the multiple losses we proposed, including an intra-domain adversarial loss and inter-domain adversarial loss. $L_{GAN2}$ further reduced the difference in stain style between the source images and the reconstructed target images. We adopted the $L_1$-distance pixel-by-pixel loss rather than a higher-order loss to encourage less blur. The style reconstruction G can be further refined by the task network loss $L_{Task}$.

The performance of the proposed model may be affected by the combination of the various loss functions and their hyperparameters. Additionally, the gradients calculated with different loss terms consume significant amounts of GPU memory, thus reducing the training speed. In the stain removal module, we encoded the color mask according to the value of each pixel's three channels in the RGB color space. When the three channels have similar values, the color mask cannot reflect whether the pixel belongs to red or blue.

In future work, we will apply the proposed framework to more datasets, including histopathological images, and more tasks including cell segmentation and detection to verify its general applicability and potential.

## ACKNOWLEDGMENT

We thank the Optical Bioimaging Core Facility of WNLO HUST for the support in data acquisition.